\documentclass[amsmath,amssymb,nofootinbib,prd]{revtex4}
\pdfoutput=1
\usepackage{graphicx}
\usepackage{amsmath}
\usepackage{bm}
\renewcommand{\mathbf}{\bm}
\newcommand{\be}{\begin{eqnarray}}
\newcommand{\ee}{\end{eqnarray}}
\def\D{\mbox{D}\,}

\begin{document}

\title{The cosmological background of vector modes}

\author{Teresa Hui-Ching Lu$^1$}
\email{teresa.huichinglu@gmail.com}

\author{Kishore Ananda$^1$}
%\email{kishore.ananda@gmail.com}

\author{Chris Clarkson$^1$}
%\email{chris.clarkson@uct.ac.za}

\author{Roy Maartens$^2$}
%\email{roy.maartens@port.ac.uk}

\affiliation{$^1$ Cosmology \& Gravity Group, Department of
Mathematics and Applied Mathematics, University of Cape Town,
Rondebosch 7701,
Cape Town, South Africa\\
$^2$ Institute of Cosmology and Gravitation, University of
Portsmouth, Dennis Sciama Building, Burnaby Road, Portsmouth PO1
3FX, UK}

\date{\today}

\begin{abstract}

We investigate the spectrum of vector modes today which is
generated at second order by density perturbations. The vector
mode background that is generated by structure formation is small
but in principle it contributes to the integrated Sachs-Wolfe
effect, to redshift-space distortions and to weak lensing. We
recover, clarify and extend previous results, and explain
carefully why no vorticity is generated in the fluid at second
order. The amplitude of the induced vector mode in the metric is
around 1\% that of the first-order scalars on small scales.  We
also calculate the power spectrum and the energy density of the
vector part of the shear at second order.
%On sub-Hubble scales it is
%insignificant but on super-Hubble scales it eventually becomes larger than the
%first-order shear.

\end{abstract}

\maketitle

\section{INTRODUCTION}

Cosmological vector perturbations at linear order satisfy an
evolution equation and a momentum constraint equation which can be
found by calculating the $i-j$ and $0-i$ parts of the Einstein
field equations respectively. In the case of a perfect fluid (or
scalar field) there is no source in the vector evolution equation
and its solutions decay as $1/a^2$. In the standard models of
structure formation, inflation does not generate vector
perturbations, and we can therefore ignore vector perturbations at
first order. However, vector perturbations will be generated at
nonlinear order by the growth of density perturbations.  Here we
revisit the analysis of this cosmological vector background (see
Refs.~\cite{Matarrese:1997ay,Mollerach:1997up,Noh:2003yg,
Mollerach:2003nq,Tomita:2005et,Mena:2007ve,Lu:2007cj} for previous
work). As we show below, the amplitude of the induced vectors is
surprisingly large on small scales, when compared to the amplitude
of the first-order Newtonian potential. In order to investigate
whether this is important dynamically we calculate the vector part
of the shear and compare it to the usual first-order scalar shear.

We consider perturbations of a flat Robertson-Walker background up
to second order: ${g}_{\mu \nu} = \bar{g}_{\mu \nu} + \delta^{(1)}
g_{\mu \nu} + \delta^{(2)} g_{\mu \nu}$. At first order,
$\delta^{(1)} g_{\mu \nu}$ only contains scalar perturbations: we
neglect the tensor perturbations, and vector perturbations are not
generated at first order in the standard model. For the
second-order perturbations, $\delta^{(2)} g_{\mu \nu}$, we project
out the vector modes. In the Poisson gauge~\cite{Matarrese:1997ay}
\begin{equation}
ds^2 = -a^2 \left[ 1 + 2 \Phi_{(1)} \right] d\eta^2 - a^2
S_{i}^{(2)}dx^id\eta + a^2  \left [1 - 2 \Phi_{(1)} \right ]
d\vec{x}\,^2, \label{perturbed metric}
\end{equation}
where $\Phi_{(1)}$ is the first-order Newtonian potential (we
assume zero scalar anisotropic stress at first order), and
$S^{i}_{(2)}$ describes the gauge-invariant second-order vector
modes, so that $\partial_{i} S_{(2)}^{i} = 0$. We effectively
ignore the second-order scalar mode $\Phi_{(2)}$ since we are
interested only in the second-order vector modes: second-order
modes can be consistently split into scalar and vector, but the
equations for the second-order vector modes will contain source
terms that are quadratic in the first-order scalar modes. In what
follows we will drop the order indices when there is no ambiguity.
With this in mind, the fluid four-velocity is given by
 \be
u_\mu=a \left[-1-\Phi+{1\over
2}\Phi^2-{1\over2}v_{(1)}^jv_j^{(1)}\,,~~ v^{(1)}_i+ {1\over2}
\left\{ v^{(2)}_i-S_i\right\} -2\Phi v^{(1)}_i \right],
\label{4vel}
 \ee
where $v^{(1)}_i=\partial_iv_{(1)}$ and $\partial_i v_{(2)}^i=0$.

The background dynamics are given by
 \be
\mathcal{H}^2 = \frac{8}{3} \pi G a^2 \rho + \frac{1}{3} a^2
\Lambda\,,~~~  \mathcal{H}' = -4\pi G (1 + w)a^2\rho +
\mathcal{H}^2\,,\label{bg}
 \ee
where $w=p/\rho=\,$const (later we specialize to $w=0$). The
first-order perturbed field equations lead to~\cite{Kodama}
 \be
v_{(1)} &=& - {
 \Phi' + \mathcal{H}  \Phi \over 4\pi G \rho a^2 (1 + w)}\,,
\label{first order velocity}\\
\delta \rho &=& {{\nabla^2} \Phi-{3 \mathcal{H}}( \Phi'  +{\cal
H}\Phi)
 \over 4\pi Ga^2} \,, \label{first order rho}
 \ee
where the Newtonian potential obeys the evolution equation
\begin{equation}
\Phi''+ 3 \mathcal{H} (1 + w) \Phi' +  \left [ (1 + w)\Lambda a^2
- w \nabla^2 \right] \Phi= 0\,. \label{equation of motion for
Bardeen potential}
\end{equation}

All linear scalar modes are determined by $\Phi$, and terms
quadratic in $\Phi$ and its derivatives will source the
second-order vector modes. Therefore we will require the power
spectrum ${\cal P}_\Phi$. In the matter era ($w=0$), neglecting
$\Lambda$, we have $a=a_0(\eta/\eta_0)^2$, and the solution in
Fourier space is $\Phi_m({k},\eta)=A_m({k})$, where we remove the
decaying mode. Then the power spectrum today is computed using the
growth suppression and transfer functions. (See Appendix~A for
more details.)

Vector perturbations typically produce vorticity and a transverse
shear in the fluid four-velocity, Eq.~\eqref{4vel}. In order to
compute them, we need the covariant definitions
(see~\cite{Tsagas:2007yx} for a recent review):
 \be
{\omega}_{\mu\nu} &=& \delta^{(1)} \omega_{\mu\nu} + \delta^{(2)}
\omega_{\mu\nu} = {h}^{\alpha}_{[\mu} {h}^{\beta}_{\nu]}
{u}_{\alpha;\beta}\,,\label{vor}\\ {\sigma}_{\mu\nu} &=&
\delta^{(1)} \sigma_{\mu\nu} + \delta^{(2)} \sigma_{\mu\nu} =
\left\{ {h}^{\alpha}_{(\mu} {h}_{\nu)}^{\beta} - \frac{1}{3}
{h}_{\mu\nu} {h}^{\alpha \beta} \right\}{u}_{\alpha;\beta}\,,
\label{shear}
 \ee
where $h_{\mu\nu}=g_{\mu\nu}+u_\mu u_\nu$ is the projector into
the instantaneous fluid rest space, and ${\omega}_{\mu\nu}u^\nu
=0= {\sigma}_{\mu\nu}u^\nu $. The vorticity is a purely vector
quantity. The vector part of the second order shear is defined via
(see Appendix D for how to get $\sigma_{j}$ from $\delta^{(2)}
\sigma_{ij}$)
 \be \label{vshear}
\delta^{(2)} \sigma_{ij}=a\partial_{(i}\sigma_{j)}\,, ~~~
\partial_i\sigma^i=0\,.
 \ee
At first order, there is only scalar shear:
 \be
\delta^{(1)} \omega_{ij} &=& 0 \,, \\ \delta^{(1)} \sigma_{ij}& =&
a\left(\partial_i \partial_j   - \frac{1}{3}{\delta}_{ij}
\nabla^2\right) v_{(1)} \,. \label{first order shear with
velocity}
 \ee

\section{SECOND ORDER VECTOR MODES}

\subsection{Vorticity}

We use a covariant and fully nonlinear approach in this
sub-section, which is more direct and transparent than a
perturbative approach in the case of vorticity, and which also
leads to a more general result.

The vorticity tensor of a fluid with four-velocity $u^\mu$ defines
the vorticity vector $\omega_\mu=\varepsilon_{\mu\nu\gamma}
\omega^{\nu\gamma}/2$, where $\varepsilon_{\mu\nu\gamma}$ is the
covariant permutation tensor in the fluid rest space. The
vorticity vector obeys the following propagation equation, which
is covariant and allows for full
nonlinearity~\cite{Tsagas:2007yx}:
 \be \label{vorp}
h_\mu{}^\nu \dot {\omega}_\nu=-{2\over3}\Theta\omega_\mu
-{1\over2} \mbox{curl}\,\dot{u}_\mu+\sigma_{\mu\nu}\omega^\nu\,.
 \ee
Here $\Theta=\nabla_\mu u^\mu$ is the volume expansion rate, an
overdot denotes covariant differentiation along the fluid flow
($u^\mu\nabla_\mu$), so that $\dot{u}_\mu$ is the fluid
four-acceleration, and the covariant spatial curl is defined by
curl$\,n_\mu=\varepsilon_{\mu\nu\gamma}{\rm D}^\nu n^\gamma$,
where ${\rm D}_\mu$ is the spatially projected covariant
derivative (${\rm D}_\mu n_\nu=h_\mu{}^\alpha h_\nu{}^\beta
\nabla_\alpha n_\beta $).

In order to evaluate the curl of the acceleration, we need the
momentum conservation equation~\cite{Tsagas:2007yx},
 \be \label{mom}
h_\mu{}^\nu \dot{q}_\nu+{4\over3}\Theta q_\mu+(\rho+p)\dot{u}_\mu+
{\rm D}_\mu p+ {\rm D}^\nu\pi_{\mu\nu}+ \sigma_{\mu\nu}q^\nu-
\varepsilon_{\mu\nu\gamma}\omega^\nu q^\gamma - \pi_{\mu\nu}
\dot{u}^\nu =0\,,
 \ee
where $q_\mu$ is the spatial momentum density flux relative to
$u^\mu$ and $\pi_{\mu\nu}$ is the fluid anisotropic stress
(spatial, tracefree and symmetric). In general, Eq.~\eqref{mom}
shows that the curl of acceleration can introduce source terms for
vorticity in Eq.~\eqref{vorp}. In the case of a perfect fluid
however, we have $q_\mu=0 = \pi_{\mu\nu}$ and ${\rm D}_\mu p=c_s^2
{\rm D}_\mu\rho$, where $c_s$ is the adiabatic sound speed. Then
using the exact identity~\cite{Tsagas:2007yx}
 \be
\mbox{curl}\,{\rm D}_\mu f = -2 \dot f \omega_\mu\,,
 \ee
the curl of Eq.~\eqref{mom} gives
$(\rho+p)\mbox{curl}\,\dot{u}_\mu= 2\dot{p}\,\omega_\mu
+\varepsilon_{\mu\nu\gamma}\dot{u}^\nu {\rm D}^\gamma (\rho+p)\,.$
Using Eq.~\eqref{mom} again, we find that the second term is
proportional to $\varepsilon_{\mu\nu\gamma}{\rm D}^\nu p {\rm
D}^\gamma \rho\,$, which vanishes, since ${\rm D}_\mu p$ is
parallel to ${\rm D}_\mu \rho$ for a perfect fluid. Collecting
results, we arrive at the fully nonlinear vorticity propagation
equation for a perfect fluid:
 \be \label{vprop}
h_\mu{}^\nu \dot{\omega}_\nu +\left( {2\over 3}- c_s^2\right)
\Theta \omega_\mu-\sigma_{\mu\nu}\omega^\nu=0\,.
 \ee
(Note that the last term on the left is at least third order.)
This equation shows that there is no source for vorticity, so that
vorticity cannot be generated in a perfect fluid, at any
perturbative order.

In particular, {\em there is no generation of vorticity at
non-linear order by first-order scalar perturbations, in the case
of a perfect fluid,} and thus
 \be \label{zerov}
\delta^{(2)} \omega_{ij} = 0\,.
 \ee
If there is primordial vorticity, then it must be introduced as an
initial condition. Any primordial vorticity will simply redshift
away as the universe expands, according to Eq.~\eqref{vprop}, and
will be entirely unaffected by the growth of density
perturbations. Effectively, the density perturbations generate
metric vector perturbations, and the fluid velocity adjusts so as
to maintain zero vorticity. This is similar to what happens with
the Harrison mechanism for magnetogenesis, where the vector modes
generated at first order by defects cannot induce vorticity in the
plasma~\cite{Hollenstein:2007kg}. For more than one perfect fluid,
vorticity non-generation applies separately to each fluid, as long
as there is no momentum exchange between the fluids. This is the
case for example with cold dark matter and baryons, which interact
only gravitationally.

Returning to the perturbative analysis, we can now use
Eq.~\eqref{zerov}, together with the Einstein equations, to
determine the vector metric perturbation $S^i$.

\subsection{Vector metric perturbations}

The second-order vorticity is given by Eqs.~(\ref{4vel}) and
(\ref{vor}):
\begin{equation}
\delta^{(2)} \omega_{ij} = \frac{a}{2} \left \{ \partial_{[j}
v^{(2)}_{ i]} - \partial_{[j} S_{i]}+ 6  \partial_{[i} \Phi
\partial_{j]} v^{(1)}   + 2
\partial_{[i} v^{(1)\prime}
\partial_{j]} v^{(1)}  \right \},
\label{second order vorticity with velocity}
\end{equation}
which is in agreement with the expression
in~\cite{Matarrese:2004kq}. Then Eq.~\eqref{zerov} becomes a
constraint on $v_{(2)}^i-S^i$. However, we have another constraint
from the second order $0i$ Einstein equation~\cite{Lu:2007cj}
 \be
6 \mathcal{H}^2 \Omega_{m} (1 + w) \left[ v^{(2)}_i-S_i\right]&=&
-\nabla^2 S_i  +8\left[ 2 \Phi' \partial_i \Phi   + \frac{2}{3 \mathcal{H}^2\Omega_m} \nabla^2 \Phi
\,
\partial_i\left( \Phi'  + \mathcal{H}
 \Phi  \right)\right]^V,
\label{second order velocity}
 \ee
where $V$ denotes schematically the vector part of the quadratic
source term which can be extracted in Fourier space as shown in
Equation~\eqref{inverse integral}. Substituting Eqs.~\eqref{first
order velocity} and \eqref{second order velocity} into
Eq.~\eqref{second order vorticity with velocity}, we obtain
\begin{equation}
\delta^{(2)} \omega_{ij} = \frac{a}{12(1 + w) \mathcal{H}^2
\Omega_m} \left \{\nabla^2 \partial_{[i} S_{j]} - \frac{16}{3(1 +
w) \mathcal{H}^2 \Omega_m} (\nabla^2 \partial_{[i} \Phi)
\partial_{j]}\left( \Phi'  + \mathcal{H} \Phi \right )
 \right \}. \label{vorticity tensor}
\end{equation}

Then Eqs.~\eqref{zerov} and \eqref{vorticity tensor} imply that
 \be \label{new solution for S}
\nabla^2 S_i={16\over 3{\cal H}^2\Omega_m(1+w)}\Big\{
\nabla^2\Phi\,
\partial_i \left( \Phi'+{\cal H}\Phi\right)\Big\}^V\,.
 \ee
This directly recovers the solution that was obtained in
Ref.~\cite{Lu:2007cj} via a Fourier-space projection of the
second-order $ij$ Einstein equation, i.e., the vector anisotropic
stress constraint,
\begin{equation}
\partial_{(i}S_{j)}' + 2 \mathcal{H} \partial_{(i}S_{j)}
= a\partial_{(i}\pi_{j)}\,. \label{vector evolution equation}
\end{equation}
Here $\pi_j$ (with $\partial_j \pi^j=0$) is the effective vector
anisotropic stress from second-order density perturbations (see
~\cite{Lu:2007cj} and Appendix B).

\begin{figure*}[th!]
\begin{center}
\includegraphics[width=0.45\textwidth]{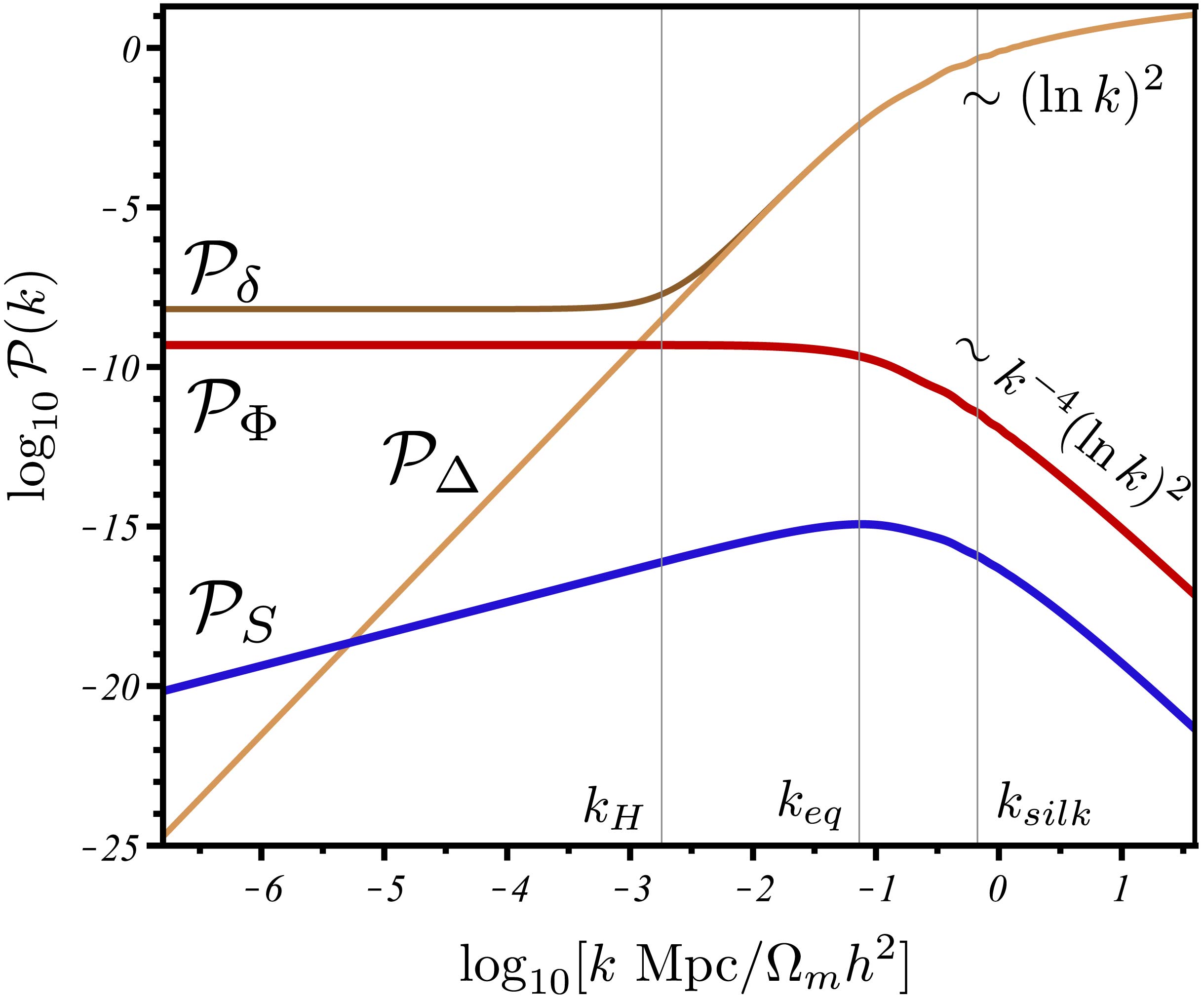}
\hfil
\includegraphics[width=0.45\textwidth]{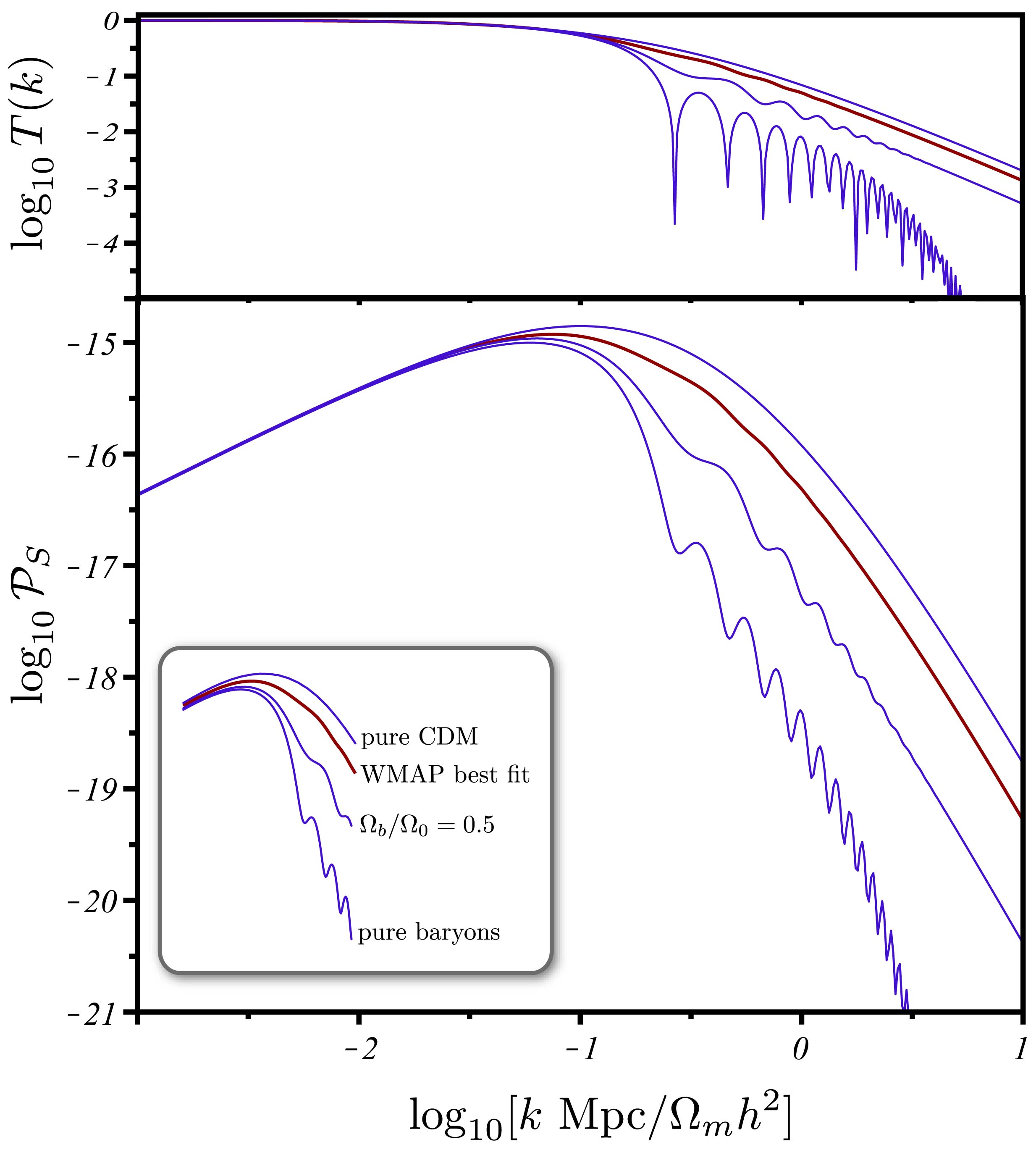}
\caption{The power spectrum today of metric vector modes generated
at second order by density perturbations. {\em Left:}
$\mathcal{P}_S$ together with the power spectra of first-order
quantities: the density perturbation $\delta=\delta\rho/\rho$ and
comoving density perturbation $\Delta=\delta-3{\cal H}v^{(1)}$,
and  $\Phi$ (using best-fit WMAP5 parameters~\cite{WMAP5}:
$\Omega_m h^2 = 0.1326, 100\Omega_b h^2=2.263, h = 0.719$). {\em
Right:} We show how increasing the baryon fraction decreases the
power in the vectors above $k_{eq}$. The baryon oscillations are
washed out to some extent in the vectors as can be seen by
comparing $\mathcal{P}_S$ with the first-order transfer function
in the top panel. }\label{power spectrum of vector modes}
\end{center}
\end{figure*}

To evaluate $S_i$ via Eq.~\eqref{new solution for S}, we work in
Fourier space, and the details are given in Appendix~\ref{PS}.
Using the first-order power spectrum (Appendix~A), the resulting
second-order vector power spectrum ${\cal P}_S $ is shown in
Fig.~\ref{power spectrum of vector modes}. This may be compared
with the power spectrum in~\cite{Lu:2007cj}, which is computed
from scalar modes in the radiation era ($w=1/3$), assuming a
power-law form for the scalar spectrum. Here we have found ${\cal
P}_S$ today, and we have calculated ${\cal P}_\Phi$ directly from
the first-order solutions, using the transfer function to relate
back to the primordial perturbations and the growth suppression
factor to take account of $\Lambda$ in the background.

In Fig.~\ref{power spectrum of vector modes} we compare the power
spectrum of the vectors with that of the Newtonian potential. For
comparison we also show the power spectrum of the two
gauge-invariant density perturbations. The amplitude of the
vectors decays on small scales, $k> k_{eq}\approx 0.009\,
\text{Mpc}^{-1}$, in contrast to the density perturbation, which
is growing, but in line with $\Phi$. For WMAP5 data~\cite{WMAP5}
 \be
{\cal P}_S\approx 6.5\times10^{-5} {\cal P}_\Phi ~~\mbox{for} ~~ k
\gtrsim k_\mathrm{silk} \approx 0.09\,\text{Mpc}^{-1} \,,
 \ee
so that the amplitude of the metric vector modes is nearly 1\%
that of the metric scalar modes on small scales. For zero baryons
we find that ${\cal P}_S\approx z_{eq}^{-1}(5.49\Omega_m
h^2-0.13)^{2.33} {\cal P}_\Phi\sim (\ln k)^2/k^4$ for $k\gtrsim
k_\mathrm{silk} \approx 0.09 \, \text{Mpc}^{-1}$. On large scales
${\cal P}_S$ scales like $k$, with a peak in the spectrum around
the equality scale. This is analogous to the peak in the induced
gravitational wave background on similar scales~\cite{BIST}. In
Fig.~\ref{power spectrum of vector modes} we also show that
including baryons induces oscillations in the vector power
spectrum (bottom panel). They are washed out in comparison to the
those present in the scalars (top panel), but are still very
prominent.

The overall shape of the $S^i$ spectrum may be understood from the
generation of vectors during the radiation era~\cite{Lu:2007cj}.
Vector modes grow outside the Hubble radius as ${a}^{1/2}$ only
through the interaction of scalar modes which are larger than the
Hubble radius. Inside the Hubble radius, vector modes decay,
slightly less rapidly than $a^{-2}$, when fluctuations in the
radiation fluid no longer support vectors. At the end of the
radiation era, vector modes with $k<k_{eq}$ have acquired a tilt
because modes are more aggressively produced by scalars which are
close to the Hubble radius~-- very long wavelength modes interact
only weakly. After equality, all vector modes grow at the same
rate, so that those which entered the Hubble radius before
equality are suppressed.

\subsection{Vector Shear}

The second order vector shear follows from Eq.~\eqref{shear} as
 \be
\delta^{(2)} \sigma_{ij}= \frac{a}{2}
\partial_{(i} v^{(2)}_{j)} + a
\Big\{\partial_{(i}v^{(1)}\partial_{j)}\left[ \Phi   +
v_{(1)}'\Big]\right\}^V= a\partial_{(i}\sigma_{j)} \,.
\label{second order shear with velocity}
 \ee
Using Eqs.~\eqref{bg}, \eqref{first order velocity} and
\eqref{second order velocity}, we have
 \be
a\partial_{(i}\sigma_{j)} & =&  \frac{a}{2} \partial_{(i} \left \{
S_{j)} - {\frac{1} {6 \mathcal{H}^2 \Omega_{m} (1 + w)}} \nabla^2
S_{j)} \right\}
 \nonumber\\&&{} -\frac{2a}{9 \mathcal{H}^4 \Omega_{m}^2
(1 + w)^2}\Big\{ \mathcal{H}^3 \left[2 + 3(2- \Omega_{m})(1 + w)
\right] \partial_i \Phi \partial_j \Phi +2 \mathcal{H}(4+3w)
\partial_i \Phi'\partial_j \Phi'  \nonumber\\&&
{}  + 2{\cal H}^2\left[
1+3(1+\Omega_m)(1+w)\right]\left(\partial_i \Phi \partial_j \Phi
\right)'-(1+2w)\nabla^2\partial_{(i} \Phi \partial_{j)}( \Phi'+
{\cal H}\Phi) \Big\}^V.
 \label{shear tensor}
 \ee

In order to find the vector power spectrum ${\cal P}_{\sigma V}$
for $\sigma_i$, we substitute Eq.~\eqref{new solution for S} into
Eq.~\eqref{shear tensor}, and then apply the vector extraction
operator in Fourier space. Further details are given in
Appendix~C. The resulting power spectrum has a similar shape to
${\cal P}_S$, since $\sigma_i=S_i/2+ \mbox{small corrections}$ by
Eq.~\eqref{shear tensor}.

It is also useful to compare the first-order scalar and
second-order vector contributions to the shear energy density. We
define the dimensionless shear density
 \be
\Omega_\sigma={a^2\over 6{\cal H}^2}\sigma_{\mu\nu}
\sigma^{\mu\nu}\,.
 \ee
The scalar and vector contributions to shear are given by
$\sigma=v_{(1)}$ [Eq.~\eqref{first order shear with velocity}] and
$\sigma_i$ [Eq.~\eqref{shear tensor}] respectively. They define
scalar power ${\cal P}_{\sigma S}$ and vector power ${\cal
P}_{\sigma V}$, which then define the spatial averages of
$\Omega_{\sigma S}, \Omega_{\sigma V} $ via
 \be
{d\langle\Omega_{\sigma S}\rangle \over d\ln k} = {a^2k^4 \over
6{\cal H}^2}\,{\cal P}_{\sigma S}\,, ~~~ {d\langle\Omega_{\sigma
V}\rangle \over d\ln k} = {a^2k^2 \over 12{\cal H}^2}\,{\cal
P}_{\sigma V}\,. \label{omsv}
 \ee
Note that there are 2 polarizations implicit in $ {\cal P}_{\sigma
V}$.

\begin{figure*}[th!]
\begin{center}
%\includegraphics[width=3in,height=2in]{power_spectrum_of_S.eps}
%\quad
\includegraphics[width=0.45\textwidth]{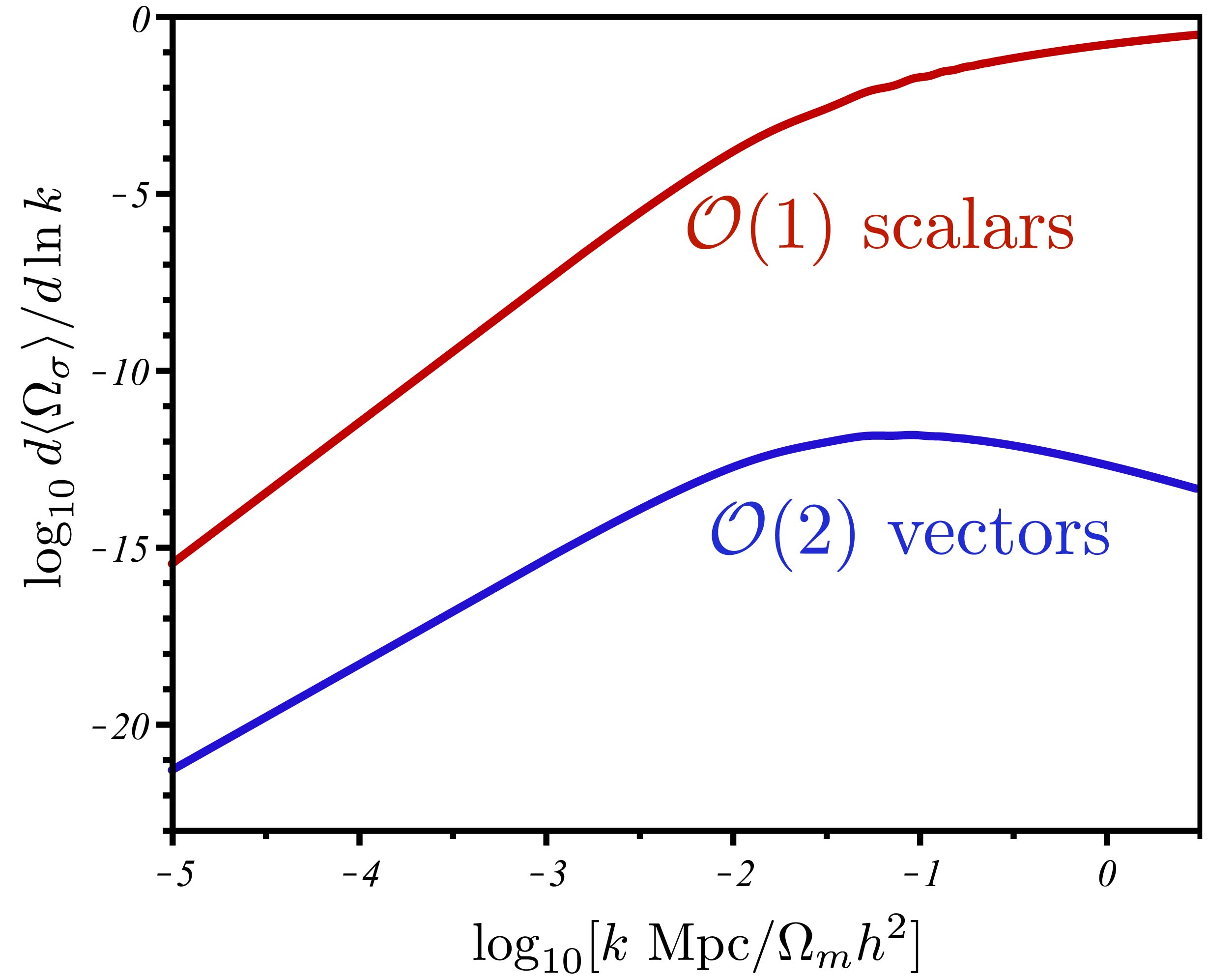}
\caption{ The energy densities of the first-order scalar and
second-order vector shear per logarithmic $k$-interval.
}\label{fig2}
\end{center}
\end{figure*}

The quantities in Eq.~\eqref{omsv} are shown in Fig.~\ref{fig2}.
As $k \rightarrow 0$, $d \langle \Omega_{\sigma{V}} \rangle/d \ln
k \sim k^3$ and $d \langle \Omega_{\sigma{S}} \rangle/d \ln k \sim
k^4$. This shows that the second-order vector shear is much
smaller than the linear scalar shear -- except on very large
scales. However, this feature in the Poisson gauge will not lead
to any growing physical effect, since both quantities are decaying
on large scales.

The vector part of the shear may be interpreted as a rotational
quantity, even though the vorticity of the fluid is zero. We can
see this readily via the covariant approach of section~IIA. The
vector shear is
 \be
\sigma_{\mu\nu}={\rm D}_{\langle\mu}\sigma_{\nu\rangle}:=\left\{
{h}^{\alpha}_{(\mu} {h}_{\nu)}^{\beta} - \frac{1}{3} {h}_{\mu\nu}
{h}^{\alpha \beta} \right\} \D_{\alpha}\sigma_{\beta}\,, ~~~ {\rm
D}_\mu\sigma^\mu=0\,.
 \ee
By boosting from the fluid rest frame to a different frame,
$\tilde u^\mu=u^\mu+v^\mu$ (with $v_\mu u^\mu=0$), the vorticity
and shear become~\cite{Tsagas:2007yx}
 \be
\tilde\omega_\mu=-\frac{1}{2}\mbox{curl}\, v_\mu\,, ~~~
\tilde\sigma_{\mu}=\sigma_{\mu}+ v_{\mu}\,.
 \ee
If we choose the frame by $v_\mu=-\sigma_\mu$, the vector part of
the shear $ \tilde\sigma_{\mu}$ is zero, but the vorticity
$\tilde\omega_\mu$ no longer vanishes. This shows the essentially
rotational nature of the fluid vector shear, even though the
vorticity of the fluid is zero. Note that it is not possible to
boost away the scalar or tensor part of the shear in this way.

\section{CONCLUSIONS}

We have computed the power spectra for the metric vector
perturbations and vector shear at second order, generated by
first-order scalar perturbations in a $\Lambda$CDM model. In
addition, we used a covariant approach to show explicitly how
vorticity is not generated in a perfect fluid at any perturbative
order by first-order scalar perturbations, so that there is no
vorticity in the matter. In order to generate vorticity, one
requires either an imperfect fluid, or momentum exchange between
the fluid and another fluid. Momentum exchange (via Compton and
Coulomb interactions) between electrons, protons and photons in
the radiation and recombination eras can generate vorticity and
magnetic fields at second
order~\cite{Matarrese:2004kq,Gopal:2004ut,Takahashi:2005nd,
Ichiki:2006cd,Siegel:2006px,Kobayashi:2007wd,Maeda:2008dv}.

In order to obtain the vector quantities $S^i$ and $v_{(2)}^i$, we
used the vanishing of vorticity [Eq.~\eqref{zerov}] and the $0i$
Einstein constraint [Eq.~\eqref{second order velocity}].
Alternatively, one could also use the $ij$ Einstein equation
[Eq.~\eqref{vector evolution equation}] and the momentum
conservation equation, which has the form
 \be \label{momentum conservation equation}
\left\{v_{(2)i}-S_{i}\right\}'+\mathcal{H}\left(1-3c_s^2\right)
\left\{ v_{(2)i}-S_{i}\right\} = M_i\,,
 \ee
where the source term $M^i$ is given in Appendix~B.

The cosmological background of vector modes is small, especially
if measured in terms of the dimensionless shear density, as shown
in Figs.~\ref{power spectrum of vector modes} and \ref{fig2}.
However, given that the amplitude of the vector modes in the
metric is as large as $\sim\,$1\% of the metric first-order scalar
modes, in principle these vector modes will have an effect on
various cosmological observations. In particular:

\begin{itemize}

\item
Redshift-space distortions~\cite{Smith:2007sb}: the divergenceless
velocity $v_{(2)}^i$ will make a contribution to radial peculiar
velocities and thus to redshift-space distortions.

\item
Large-angle CMB temperature
anisotropies~\cite{Mollerach:1997up,Tomita:2005et}: the vector
modes will contribute to the Doppler and integrated Sachs-Wolfe
effects:
 \be
\delta^{(2)}T = {1\over2}\left\{v^{(2)}_i-S_i\right\}e^i\Big|_E^O
+ {1\over2}\int_E^O d\lambda\,\partial_i S_j \,e^i e^j\,.
 \ee

\item
Weak lensing~\cite{sef,Durrer:1994uu}: vector modes produce a
deflection angle
 \be
\vec\alpha = \int_E^O d\lambda \left(\vec\nabla \times \vec S
\right)\times \vec e\,.
 \ee

\item
CMB polarization~\cite{Kamionkowski:1996zd,Mollerach:2003nq}: the
vector modes will leave a characteristic imprint on CMB
polarization.

\end{itemize}

Further work is needed to compute the size of these vector
corrections. They are likely to be significant mainly below $\sim
10\,$Mpc, but in this regime the scalar non-linear effects are
important and likely to dominate.

The vector degree of freedom forms an integral part of the
perturbative expansion when one goes beyond linear order. At the
order we have considered, vectors must be present essentially
through a constraint in the field equations arising at order
$\Phi^2$, even though vectors have no independent propagating
degrees of freedom. This is distinct from the intrinsically
propagating degree of freedom in the scalar-induced gravitational
wave background~\cite{ACW,BIST,MBMR}. We have shown that the
vector mode background has maximum power around the equality
scale, similar to the induced gravitational wave background, and
the metric vector modes achieve their maximal fraction of the
linear metric scalar modes on scales below the Silk scale. The
size of the vector contribution to the full non-linear power
spectrum relevant for structure formation remains to be
calculated.

\[ \]{\bf Acknowledgements:}\\
We thank Bruce Bassett, Marco Bruni, Ruth Durrer, Cyril Pitrou,
Jean-Phillipe Uzan and David Wands for useful discussions and
comments.  THCL, KA and CC are supported by the National Research
Foundation (South Africa), and RM is supported by STFC (UK). THCL
and CC acknowledge financial support from the University of Cape
Town. KA is additionally supported by the Italian {\it Ministero
Degli Affari Esteri-DG per la Promozione e Cooperazione Culturale}
under the joint Italy/ South Africa Science and Technology
agreement. THCL, KA and CC thank the ICG Portsmouth for
hospitality during various stages of the work; their visits were
partly supported by a joint Royal Society (UK) - National Research
Foundation (South Africa) grant UID 65329.

\appendix

\section{First-order scalar power spectrum}

The power spectrum for the first-order scalar perturbations is
defined by
\begin{equation}
\langle \Phi^{\ast}(\mathbf{k},\eta) \Phi(\mathbf{k}',\eta)
\rangle = \frac{2 \pi^2}{k^3} \delta^3(\mathbf{k} - \mathbf{k}')
\mathcal{P}_{\Phi}(k, \eta).\label{ps}
\end{equation}
In the early radiation era,
\begin{equation}
\mathcal{P}_{\Phi_{r}}(k) \approx {A}_r(k)^2
\frac{k^3}{486\pi^2}\,, ~~{A}_r(k)^2 \approx \frac{216\pi^2}{k^3}
\Delta_{\mathcal{R}}^{2}(k),
\end{equation}
where $\Delta_{\mathcal{R}}^{2}$ is the primordial power of the
curvature perturbation, with~\cite{WMAP5}
$\Delta_{\mathcal{R}}^{2} \approx 2.41 \times 10^{-9}$ at a scale
$k_{CMB}=0.002 \mathrm{Mpc}^{-1}$. By conservation of the
curvature perturbation, $\Phi_m=9\Phi_r/10$ at equality, so the
early matter power is given by
\begin{equation}
\tilde{A}_m(k) = \frac{\sqrt{3}}{30g_\infty} A_r(k) \approx
\frac{3\sqrt{2}\pi\Delta_\mathcal{R}}{5 g_\infty k^{3/2}},
\end{equation}
assuming a scale-invariant initial spectrum. Here $g_\infty$ is a
normalization parameter in the $\Lambda$ growth suppression
function
\begin{equation}\label{gfac}
g(z) = \frac{5}{2} g_{\infty} \Omega_{m}(z) \left\{
\Omega_{m}(z)^{4/7} - \Omega_\Lambda(z) + \left[ 1 + \frac{1}{2}
\Omega_{m}(z)\right] \left[1 + \frac{1}{70} \Omega_\Lambda(z)
\right] \right\}^{-1},
\end{equation}
and $g_\infty$ is chosen so that $g(0)=1$. The power today is
given by $A_m(k) = \tilde{A}_m(k) T(k)$, where $T(k)$ is the
normalized transfer function~\cite{Eisenstein} (see
Fig.~\ref{power spectrum of vector modes}): \be
\mathcal{P}_{\Phi}=\left( \frac{3 \Delta_\mathcal{R}}{5
g_{\infty}} \right )^2 g^2 T(k)^2. \ee

\section{Vector evolution and momentum conservation}

The source term in the vector evolution equation Eq.~\eqref{vector
evolution equation} is as follows (where we assume $w=\,$const):

 \be
a\partial_{(i}\pi_{j)} =  4 \left[ 1 + \frac{2}{3 \Omega_m (1 +
w)} \right] \Big\{\partial_{i} \Phi \partial_{j} \Phi\Big\}^V +
\frac{8}{3 \mathcal{H}^2 \Omega_m (1 + w)} \Big\{
\partial_{i} \Phi' \partial_{j} \Phi'  +
2 \mathcal{H}  \partial_{(i} \Phi
\partial_{j)} \Phi'  \Big\}^V, \label{EFESV without v}
 \ee
where $V$ denotes the vector part, and we have used the fact that
$\big\{ \Phi\partial_{i} \partial_{j}
\Phi\big\}^V=-\big\{\partial_{i} \Phi \partial_{j} \Phi\big\}^V$.

In Eq.~\eqref{momentum conservation equation}, the source term in
the vector momentum conservation equation is as follows (where we
also assume $w=\,$const):

\begin{eqnarray}
9\mathcal{H}^4\Omega_{m}^2 (1+w)^2 M_i&=&6\mathcal{H}^3\left[
24(1+c_s^2)^2 - 12(1+c_s^2)(2+w) + \Omega_{m}(1+w)(5+18 c_s^2
-15w) \right]\left\{\Phi\partial_{i}
\Phi'\right\}^{V}\nonumber\\
&&+6\mathcal{H}^3\left[ 24(1+c_s^2)^2 - 12(1+c_s^2)(2+w) +
2\Omega_{m}(1+w)(3c_s^2 -2)
\right]\left\{\Phi'\partial_{i}\Phi\right\}^{V}\nonumber\\
&&+2\mathcal{H}\left[ 10(1+c_s^2) - 18(1+c_s^2)^2 -
6(1+c_s^2)(1+w) -4
\right]\left\{\mathcal{H}\partial_{i}\Phi\nabla^2\Phi+\partial_{i}
\Phi'\nabla^2\Phi\right\}^{V}\nonumber\\
&&-{36\mathcal{H}^2 c_s^2}\left[ 2(1+c_s^2) +\Omega_{m}(1+w)
\right]\left\{\Phi\nabla^{2}\partial_{i}\Phi\right\}^{V} +{8
c_s^2} \left\{\mathcal{H}\nabla^{2}\Phi'\partial_{i}\Phi
+\nabla^{2}\Phi'\partial_{i}\Phi'\right\}^{V}\nonumber\\
&&+{4(1+c_s^2)c_s^2}
\left\{\nabla^{2}\Phi\nabla^{2}\partial_{i}\Phi\right\}^{V}
-24\mathcal{H}(1+c_s^2)c_s^2
\left\{\Phi'\nabla^{2}\partial_{i}\Phi\right\}^{V}\nonumber\\
&&+8\mathcal{H}
\left\{\partial^{k}\Phi'\partial_{k}\partial_{i}\Phi
+\partial^{k}\Phi\partial_{k}\partial_{i}\Phi' \right\}^{V}.
\end{eqnarray}

\section{Vector Mode Power Spectrum}\label{PS}

We define the Fourier transform of the vector perturbation as
\begin{equation}
S_i(\mathbf{x}, \eta) = \frac{1}{(2 \pi)^{3/2}} \int d^3 k \left [
S(\mathbf{k}, \eta) e_i(\mathbf{k}) + \bar{S}(\mathbf{k}, \eta)
\bar e_i(\mathbf{k}) \right ]{e}^{i \mathbf{k}\cdot\mathbf{x}},
\end{equation}
where the two orthonormal basis vectors $\mathbf{e}$ and
$\bar{\mathbf{e}}$ are orthogonal to $\mathbf{k}$ with
\begin{equation}
S(\mathbf{k}, \eta) = \frac{1}{(2 \pi)^{3/2}} \int d^3 x
S_i(\mathbf{x}, \eta) e^{i}(\mathbf{k})
{e}^{-i\mathbf{k}\cdot\mathbf{x}}. \label{inverse integral}
\end{equation}
The inverse integral is similarly defined for its second parity
$\bar{S}(\mathbf{k}, \eta)$. Equation~\eqref{new solution for S}
gives a solution for $S\left ( \mathbf{k}, \eta \right )$ in
Fourier space
\begin{equation}
S(\mathbf{k},\eta) = \frac{16i}{3 \Omega_m (2 \pi)^{3/2}}
\frac{e^j(\mathbf{k})}{k^2} \int d^3 k'\ |\mathbf{k} -
\mathbf{k}'|^2 k'_j \mathcal{B}(\mathbf{k} - \mathbf{k}',
\mathbf{k}', \eta), \label{new solution for vector mode in Fourier
space}
\end{equation}
where
\[
\mathcal{B}(\mathbf{k}_1,\mathbf{k}_2, \eta) = {\mathcal{H}^{-2}}
\Phi(\mathbf{k}_1, \eta) \left [ \Phi'(\mathbf{k}_2, \eta) +
\mathcal{H} \Phi(\mathbf{k}_2, \eta) \right ].
\]

Defining the power spectrum as
\begin{equation}
\left \langle S^{\ast} \left ( \mathbf{k}, \eta \right ) S \left (
\mathbf{k}', \eta \right ) \right \rangle = \frac{2 \pi^2}{k^3}
\delta^3 \left ( \mathbf{k} - \mathbf{k}' \right ) \mathcal{P}_S
(k, \eta). \label{power spectrum for vector mode}
\end{equation}
 we find, using Wick's theorem,
\begin{equation}
\begin{split}
\mathcal{P}_{S} (k, \eta) & = \frac{16}{9 k \Omega^2_m \pi^5} \int
d^3 k' |\mathbf{k} - \mathbf{k}'|^2 \left [k'_j e^j(\mathbf{k})
\right] \left [k'_m e^m(\mathbf{k}) \right]
\mathcal{B}(|\mathbf{k} - \mathbf{k}'|,k', \eta) \\ &
\times \left \{ |\mathbf{k} - \mathbf{k}'|^2
\mathcal{B}(|\mathbf{k} - \mathbf{k}'|, k',\eta) - (k')^2
\mathcal{B}( k',|\mathbf{k} - \mathbf{k}'|, \eta) \right \}.
\label{power spectrum of S before projection}
\end{split}
\end{equation}
This may be simplified to give
\begin{equation}\label{Pi}
\mathcal{P}_{S}(k) = \left( \frac{2 \Delta_\mathcal{R}}{5
g_{\infty}} \right )^4\left( \frac{3 {g} \left[ g' + \mathcal{H} g
\right ]}{\Omega_m\mathcal{H}^2} \right )^2 \,k^2\, \Pi(u^2),
\end{equation}
where
\begin{equation}
\Pi(\xi) = \int^{\infty}_{0} d v\ \int^{v + 1}_{|v - 1|} d u\ \xi
\ (uv)^{-2} (u^2-v^2) \left[4 v^2 -(1+v^2-u^2)^2 \right] \left[
T(kv) T(ku) \right]^2, \label{convolution function}
\end{equation}
and $v = {k'}/{k},\ u = \sqrt{1 + v^2 - 2 v \cos \theta}$ and
$\cos\theta=\mathbf{k}'\cdot\mathbf{k}/(k'k)$.

\section{Vector shear power spectrum}

To extract the divergenceless vector $\sigma_i$ from
$\partial_{(i} \sigma_{j)}$, we use the operator
${\mathcal{V}}_{m}^{ij}$~\cite{Lu:2007cj}
\begin{equation}
{\mathcal{V}}_{m}^{ij}(\mathbf{x}, \mathbf{x}') = -\frac{2 i}{(2
\pi)^3} \int d^3 k'\ {k'^{-2}} \int d^3 x'\ {k'^{i}} \left [
e_{m}(\mathbf{k}') e^{j}(\mathbf{k}') + \bar{e}_{m}(\mathbf{k}')
\bar{e}^{j}(\mathbf{k}') \right ]  e^{i
\mathbf{k}'\cdot(\mathbf{x} - \mathbf{x}')}.
\end{equation}
Then $\sigma_m(\mathbf{x}) = {\mathcal{V}}_{m}^{ij}(\mathbf{x},
\mathbf{x}') \partial_{(i} \sigma_{j)} (\mathbf{x}') $, and
Eqs.~\eqref{new solution for S} and~\eqref{shear tensor} lead to
the second-order vector shear in Fourier space
 \be
\sigma(\mathbf{k}, \eta) &=& \frac{-2ai}{3 \Omega_m (2 \pi)^{3/2}}
\frac{e^j(\mathbf{k})}{k^2} \int d^3 k' k'_j \Big \{ \left ( k^2 -
6(k'_i k^i) - 4|\mathbf{k} - \mathbf{k}'|^2 \right )
\mathcal{B}(\mathbf{k} - \mathbf{k}', \mathbf{k}',\eta)
\nonumber \\
&&  {} + \frac{2}{3 \Omega_m} \left ( k^2 - 2k'_i k^i \right)
\left [ \mathcal{C}(\mathbf{k}',\mathbf{k} - \mathbf{k}', \eta) +
\left( 1 - \frac{3 \Omega_m}{2} \right )
\mathcal{B}(\mathbf{k}',\mathbf{k} - \mathbf{k}', \eta) \right ]
\Big \}, \label{vector shear in Fourier}
 \ee
where
 \be
\mathcal{C}(\mathbf{k}_1,\mathbf{k}_2, \eta) =
\frac{1}{\mathcal{H}^3} \Phi'(\mathbf{k}_1, \eta) \left [
\Phi'(\mathbf{k}_2, \eta) + \mathcal{H} \Phi(\mathbf{k}_2, \eta)
\right ].
 \ee
The power spectrum is defined as in Eq.~\eqref{power spectrum for
vector mode}. By Eq.~\eqref{vector shear in Fourier} and Wick's
theorem, we obtain
 \be
\mathcal{P}_{\sigma V} (k, \eta) & =&  \left(
\frac{\Delta_\mathcal{R}}{g_{\infty}} \right )^4\left[ \frac{3 k a
(g' + \mathcal{H} g)}{50 \Omega_m \mathcal{H}^2} \right ]^2 \Big
\{-2 g^2 \Pi(2 + u^2 + 3v^2) - \frac{8}{3 \mathcal{H} \Omega_m} g
\nonumber \\ &&{}
 \times \left[ g' + \left( 1 - \frac{3 \Omega_m}{2} \right )
\mathcal{H} g \right] \Pi(1 + 2 v^2) + \frac{8}{9 \mathcal{H}^2
\Omega_m^2} \left[ g' + \left( 1 - \frac{3 \Omega_m}{2} \right )
\mathcal{H} g \right]^2 \Pi(u^2-v^2) \Big\},
 \ee
where $g$ is given by Eq.~\eqref{gfac} and $\Pi$ by
Eq.~\eqref{convolution function}.

%\bibliography{Lu}

\end{document}